\documentstyle[aps,12pt,epsfig]{revtex}

\begin{document}
\title{Proton-proton Cross Section
at $\sqrt{s}\sim 30$ TeV\thanks{Research supported in part
by the U.S. Department of Energy under Grant Number  DE FG02 01
ER 40626}\\[5mm]}

\author{Ralph Engel$^1$, T.K. Gaisser$^1$, 
Paolo Lipari$^2$ \& Todor Stanev$^1$}
\address{$^1$Bartol Research Institute\\
University of Delaware, Newark, DE 19716\\
$^2$ Dipt. di Fisica and INFN\\
Universit'{a} di Roma I, Piazzale A. Moro 2\\
00185 Roma, Italy}

\maketitle

\begin{abstract}
There are both theoretical and experimental uncertainties in
using data from cosmic--ray air showers to estimate hadronic
cross sections above accelerator energies. 
 We outline these problems and compare the physics used to 
 extract $\sigma_{pp}^{\rm tot}$ from air shower data to the
 widely used parameterization of the proton--proton
 cross section of Donnachie and Landshoff \cite{Donnachie92b}
 and other contemporary models. We conclude that the published
 cosmic--ray cross section values do not strongly constrain
 $\sigma_{pp}^{\rm tot}$ fits from lower energy accelerator data.   
\end{abstract}

\section*{Introduction}

New and proposed experiments to study the cosmic-ray spectrum up to
$10^{20}$~eV and beyond
\cite{Abu-Zayyad97a,Boratav92a,Teshima97a,Ormes96a,Linsley97a,Linsley97b}
will depend for their interpretation on extrapolations of models of
hadronic interactions more than two orders of magnitude in center
of mass energy beyond what is accessible with present colliders.
The interaction lengths of hadrons in the atmosphere, and hence
their cross sections, are the most obvious determining factor
in the rate at which the showers develop.  An extra source of model
dependence is the relation between hadron cross sections
in air and the more basic
 hadron--hadron cross sections.

Cosmic-ray measurements have been used in the past to 
determine $\sigma_{p-air}^{\rm inel}$ and, with the
help of Glauber multiple scattering theory \cite{Glauber70a},
to estimate $\sigma_{pp}^{\rm tot}$.  Frequently quoted examples
are the Fly's Eye experiment \cite{Baltrusaitis84,Baltrusaitis85} and the
Akeno experiment \cite{Honda93}.  Both experiments find
rather large central values of $\sigma_{p-air}^{\rm inel}$
($\approx 540$~mb \cite{Baltrusaitis84} and $\approx 570$~mb
\cite{Honda93} at lab energy $E_0\, \sim$ 4$\times$10$^8$ GeV).  
 In both experiments, the proton--air cross section
has to be inferred from some measure of the attenuation of
the rate of showers deep in the atmosphere.    The measured
attenuation depends on the cross section which determines
the depth at which showers are initiated, but it also depends
very significantly 
on the rate at which energy is dissipated in the subsequent atmospheric
cascades.  For this reason, a simulation which includes a full
representation of the hadronic interactions in the cascade is needed.
Because these two experiments measure the attenuation in quite different
ways, the fact that their inferred values of $\sigma_{p-air}^{\rm inel}$
agree is a non-trivial result.

Having determined $\sigma_{p-air}^{\rm inel}$, the experimental
groups go on to derive corresponding values
for $\sigma_{pp}^{\rm tot}$ of $120$~mb \cite{Baltrusaitis84,Baltrusaitis85}
and $125$~mb \cite{Honda93} 
at $\sqrt{s}$  about 30 TeV.  As noted in the Review of Particle Physics
\cite{PDG96}, $\sigma_{pp}^{\rm tot}\sim 120$~mb is  in good agreement
with extrapolation of the parameterization of Donnachie and
Landshoff (DL) \cite{Donnachie92b}.  As we discuss in the next section,
however, the cosmic--ray values of $\sigma_{pp}^{\rm tot}$ are
 based on a parameterization of the nucleon--nucleon scattering amplitude
 that is in disagreement with high energy collider data.  Therefore, the
 quoted values cannot be used
 to pin down a high energy extrapolation of the {\it pp} cross section. 

Indeed, it has been pointed out in the past \cite{Gaisser87a,Nikolaev93b}
that such large values of $\sigma_{p-air}^{\rm inel}$ ($\sim 550$~mb)
would require significantly larger values of $\sigma_{pp}^{\rm tot}$
than that predicted by the parameterization of Ref. \cite{Donnachie92b}.  
Conversely, if that predicted behavior of the hadronic cross
section is correct, then the hadron--air cross sections should
be smaller, and this could have important consequences for development
of high energy cascades.

The plan of the paper is as follows.  In the next section we discuss
the relation between the nucleon-nucleon cross section and
the nucleon-nucleus cross section, in particular, how it depends
on the slope of the elastic $pp$ cross section.  Next we review
how the hadron--air cross sections are inferred from air shower
experiments and discuss the resulting uncertainties in
$\sigma_{p-air}^{\rm inel}$
and their implications for $\sigma_{pp}^{\rm tot}$.

\section*{Proton-proton vs. proton--air cross section}

The relation
between the hadron-nucleon cross section and the corresponding
hadron-nucleus cross section depends significantly
on the elastic slope parameter $B(s)$
\begin{equation}
B(s) = {d\over dt}\left[\ln\left({d\sigma_{pp}^{\rm el}\over dt}
\right)\right]_{t=0}\ .
\label{slope-def}
\end{equation}
This relation is discussed in the context of cosmic--ray cascades in
detail in Ref.~\cite{Gaisser87a}.  Qualitatively, the relation is such that for
a given value of $\sigma_{pp}^{\rm tot}$, a larger value of the slope
parameter corresponds to a larger proton--air cross section.
Conversely for a given value of
$\sigma_{p-air}^{\rm inel}$, a larger value of $B(s)$ leads to a smaller
value of $\sigma_{pp}^{\rm tot}$.  In addition, the smaller the slope
parameter, the larger is the uncertainty in the derived proton--proton cross
section.

{}For example, the Fly's Eye value of $\sigma_{pp}^{\rm tot}=122\pm 11$~mb
at $\sqrt{s}=30$~TeV \cite{Baltrusaitis84,Baltrusaitis85} is obtained
using a outdated geometrical scaling fit \cite{DiasdeDeus78a,Buras74a} 
to extrapolate the
slope parameter to this energy.  This results in a large value
of $B>30$~GeV$^{-2}$ and hence (for a measured value of
$\sigma_{p-air}^{\rm inel}\approx 540 \pm 50$~mb) a small value of
$\sigma_{pp}^{\rm tot}$.  
Using a different model for the
slope parameter \cite{Chou68,Bourrely84a}, for example, as advocated in
the review article of Block and Cahn \cite{Block85}, leads to a
slower increase in $B(s)$ and to a considerably larger value of
$\sigma_{pp}^{\rm tot}\approx 175_{-30}^{+40}$~mb \cite{Gaisser87a}.
The same applies to the Akeno analysis
and numbers~\cite{Honda93}.

Before discussing the slope parameter further, it is useful to review
briefly the basis of the very successful DL fits of cross sections,
which are based on a one-pomeron exchange model (e.g.\
\cite{Donnachie83} and Refs. therein).  In such a model,
the energy dependence of the 
total cross section for $AB$ scattering is given by \cite{Donnachie92b}
\begin{equation}
\sigma^{\rm tot}_{AB}(s) = X_{AB} \left(\frac{s}{s_0}\right)^\Delta
+ Y_{AB} \left(\frac{s}{s_0}\right)^{-\epsilon} \ .
\label{DL-par}
\end{equation}
The constants $X_{AB}$ and $Y_{AB}$ are target and projectile specific
whereas the effective powers
$\Delta\approx 0.08$ and $\epsilon\approx 0.45$
are independent of the considered particles $A$ and $B$.
Within the uncertainties of the measurements, this parameterization
is in agreement with almost all currently available data on $pp$, $p\bar p$,
$\pi p$, $\gamma p$, and $\gamma\gamma$ total cross sections.
It should be noted that the high energy $p\bar p$ data are not fully
self-consistent. There is some disagreement between measurements 
of the total cross section at
$\sqrt{s} = 1800$ GeV. Whereas the E710 \cite{Amos90a}
and the preliminary E811 \cite{Avila97a} data
are in perfect agreement with the DL prediction
\cite{Donnachie92b}, the CDF measurement \cite{Abe94d}
shows a steeper rise of the total $p\bar p$ cross section. 
New data from HERA ($\sigma^{\rm tot}_{\gamma p}$, \cite{Aid95b})
and LEP2 ($\sigma^{\rm tot}_{\gamma \gamma}$, \cite{Acciarri97a}),
although being compatible with an energy dependence of $\Delta \approx
0.08$, indicate that the cross section may rise
faster with energy than assumed in the DL fit. 
{}Furthermore, in a recent fit to $pp$ and $p\bar p$ data \cite{Cudell97a} 
a slightly higher value of $\Delta = 0.096_{-0.009}^{+0.012}$ was found.

Given the success of the one-pomeron exchange model in predicting the
total cross section, one might apply it
to derive further predictions. The one-pomeron amplitude can be
written as 
\begin{equation}
{\cal A}(s,t) = g_{AB}(t) \left(\frac{s}{s_0}\right)^{\alpha(t)}
\label{ampl-elast}
\end{equation} 
with $\alpha(t=0) = 1+\Delta$. Collider data
on elastic scattering suggest for small $|t|$ the
functional dependence $g_{AB}(t) = X_{AB} \exp\{\frac{1}{2} B_0 t\}$.
{}Following the predictions of Regge theory, $B_0$ is an energy-independent
constant. Consequently, the elastic slope $B(s)$ is given by
\begin{equation}
B(s) = B_0 + 2 \alpha^\prime(0) \ln \left(\frac{s}{s_0}\right)
\label{slope}
\end{equation}
where the parameter $\alpha^\prime(0)$ is a constant and has to be
determined from data \cite{Donnachie83}. 
The elastic cross section follows from
\begin{equation}
\sigma^{\rm el}_{AB} = (1+\rho^2) \frac{(\sigma^{\rm tot}_{AB})^2}{16
\pi B(s)}\ .
\label{sig-el}
\end{equation}
At high energies the ratio $\rho$ between the real and the imaginary part of
the forward scattering amplitude is small and $\rho^2$ can be neglected.

In a model with geometrical scaling it is assumed that the increase of
the total cross section stems entirely from an increase of the 
transverse size of the scattering particles. 
The opacity of the particles is considered as constant. A direct
consequence of this assumption is the energy independence of the ratio
$R\,=\,\sigma^{\rm el}_{pp} (s) / \sigma^{\rm tot}_{pp} (s)$, which, in
combination with Eq.~(\ref{sig-el}), leads to the relation
\begin{equation}
B(s) = (1+\rho^2) \frac{\sigma_{pp}^{\rm tot}(s)}{16 \pi\; R}\ .
\label{geom-def}
\end{equation}
Over the ISR energy range $R\approx 0.17$, which was the value
used in (\ref{geom-def}) in Refs.~\cite{Baltrusaitis84,Honda93}.

\begin{figure}[!hbt]
\centerline{\psfig{figure=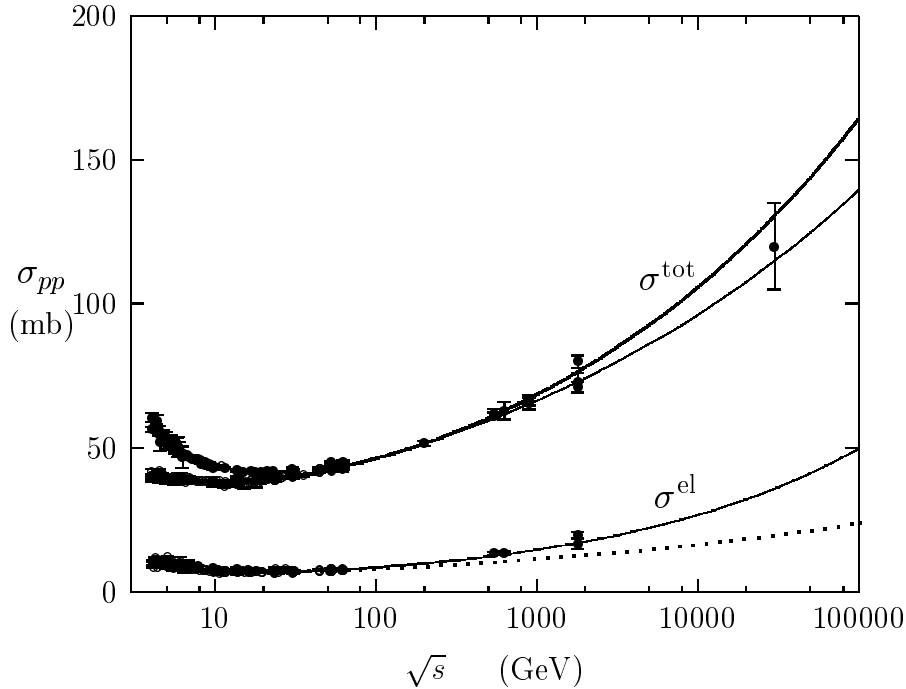,width=12cm}}
\vspace*{5mm}
\caption{\em
Data on $pp$ and $p\bar p$ interactions~\protect\cite{PDG96} 
are compared with the 
DL parameterization \protect\cite{Donnachie92b} (lower curve) and the fit
of Ref.~\protect\cite{Cudell97a} (upper curve).
The predictions for the elastic cross section from
Eq.~(\protect\ref{sig-el})
and in  the case of geometrical scaling (dotted curve) are also shown.
The data point at $\sqrt{s} = 30$
TeV is the original Fly's Eye estimate \protect\cite{Baltrusaitis84}. 
\label{pptot}}
\end{figure}
\begin{figure}[!hbt]
\centerline{\psfig{figure=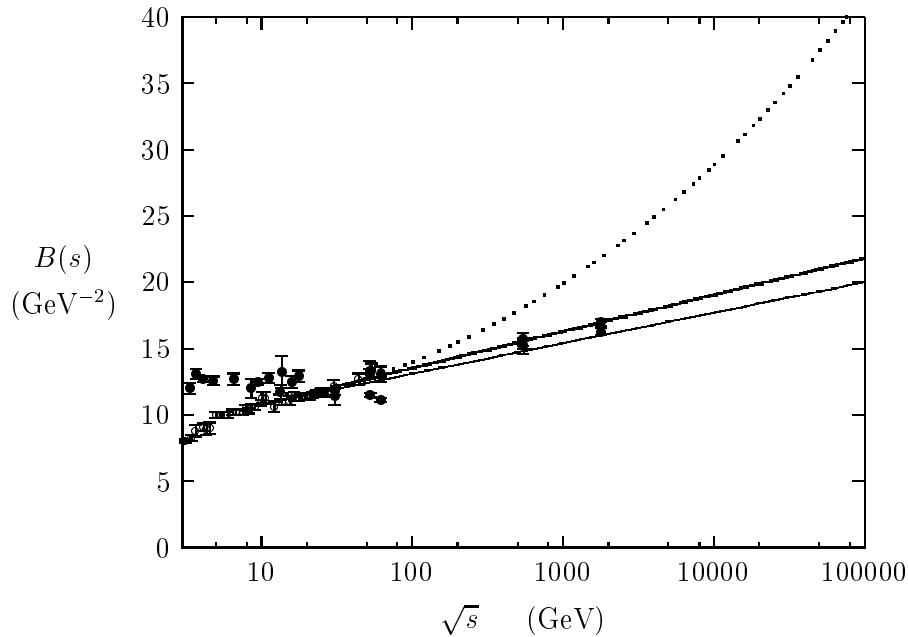,width=12cm}}
\vspace*{5mm}
\caption{\em
Elastic slope parameter for $pp$ and $p\bar p$ interactions.
The solid lines are the predictions of the one-pomeron exchange model with
$\alpha^\prime(0) = 0.25$ and $0.3$ GeV$^{-2}$. The dotted line
corresponds to geometrical scaling. The data are taken from
Refs.~\protect\cite{Castaldi85,Amos90b,Abe94b}.
\label{ppsl}}
\end{figure}

In Fig.~\ref{pptot} the parameterizations of Refs.~\cite{Donnachie92b} and
\cite{Cudell97a} are compared to data. The data point at $\sqrt{s} = 30$
TeV is the original Fly's Eye estimate \cite{Baltrusaitis84}. 
The prediction for geometrical scaling has been calculated using the
DL model for the total cross section.
Whereas both Regge
parameterizations are in agreement with data on total as well as elastic
cross sections, the geometrical scaling
model fails to describe the elastic scattering data.
This becomes even more obvious if one considers the predictions for the
energy dependence of the elastic slope parameter as shown in Fig.~\ref{ppsl}.
In contrast, the single pomeron exchange model
is in very good agreement with collider data.  Such an
$a+b\ln(s)$ extrapolation of the slope parameter is often used to fit 
data (for example, \cite{Goulianos83}) and also to estimate
cross sections and interaction lengths for cascade 
calculations \cite{Kalmykov92e,Ranft95a}.
Remarkably, the minijet calculation of Block, Halzen and Margolis
\cite{Block92} (BHM) predicts a slope parameter that almost coincides with the
one-pomeron model extrapolation using $\alpha^\prime(0) = 0.3$
GeV$^{-2}$.
As recognized by DL the single pomeron exchange model is not
consistent with unitarity.  One way to see this is to note
from Eqs.~(\ref{DL-par},\ref{slope},\ref{sig-el}), that at asymptotically
high energy the unitarity requirement\begin{equation}
{\sigma_{AB}^{\rm el}\over\sigma_{AB}^{\rm tot}}\;<\;{1\over 2}
\label{unitarity}
\end{equation}
is violated. We point out, however, that the model of BHM~\cite{Block92} 
does satisfy unitarity and it gives a similar prediction to the
single pomeron fit over the energy range shown in Fig.~\ref{ppsl}.

\begin{figure}[!hbt]
\centerline{\psfig{figure=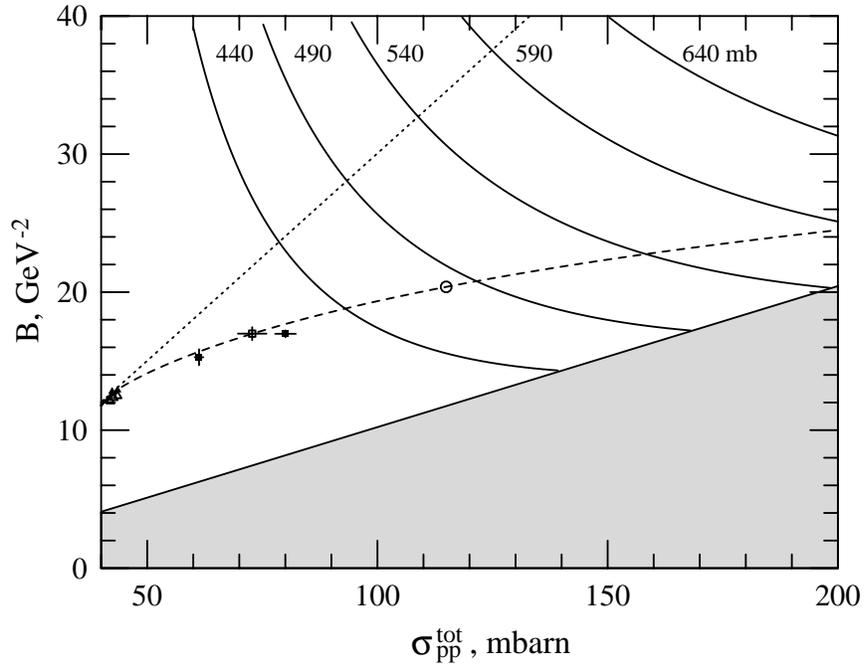,width=12cm}}
\caption{\em
 $B$ dependence on $\sigma_{pp}^{\rm tot}$ and the values of
 $\sigma_{pp}^{\rm tot}$ allowed by the Fly's Eye measurement. The shaded
 area is excluded by the unitarity constraint. Solid symbols give
 experimental data points. Dashed line shows $B$ as in DL fit; dotted line
 shows geometrical scaling. The open point indicates
 $\sigma_{pp}^{\rm tot}$ at $\protect\sqrt{s}$ = 30 TeV from the DL fit.
 The five curved lines show the region allowed by  $\sigma_{p-air}^{\rm prod}$
 = 540 mb $\pm$1$\sigma$ and $\pm$2$\sigma$ (see text).  
\label{PLfig}}
\end{figure}

 We summarize some of the results of  this section 
 in Fig.~\ref{PLfig}, by displaying them in 
the ($\sigma_{pp}^{\rm tot}$--$B$) plane. 
 The shaded region  corresponds to the region excluded by the unitarity
 constraint of Eq.~(\ref{unitarity}).  The  points  represent  experimental
 measurements at ISR (triangles) and ${\bar{p}p}$~collider (squares).
 The dotted  line indicates the  relation  between  $B$ and
 $\sigma_{pp}^{\rm tot}$ predicted by geometrical  scaling with $R=0.17$ in
 Eq.~(\ref{geom-def}). This  line fails to  describe the highest energy
 measurements. The  dashed line   corresponds to the DL fit
 to  $\sigma_{pp}^{\rm tot}$  together with 
equation (\ref{slope}) for the energy
 dependence of the slope (with $\alpha^\prime(0) = 0.3$~GeV$^{-2}$).
 Each point on the dashed line corresponds to a value of the center of mass
 energy of the  $pp$ (or $p\bar{p}$) 
reaction. We have  indicated  with a  circle
 the  point for $\sqrt{s} = 30$~TeV.

 Using the Glauber formalism a  fixed  value of the $p$--air cross 
 section  can be represented as a curve in the ($\sigma_{pp}^{\rm tot}$--$B$)
 plane. The five curved  lines in Fig.~\ref{PLfig} indicate the set of 
 values of $\sigma_{pp}^{\rm tot}$ and $B$  that result in a
 proton--air  cross section of $\sigma_{p-air}^{\rm inel}$ of 540, $540\pm 50$
 and $540 \pm 100$ mb, that is the central value and  $\pm 1, 2$~standard
 deviations  of the Fly's Eye  measurement at $\sqrt{s} = 30$~TeV.
 The  intersections  of the curves  corresponding to 590 and 490 mb
 with the dotted  line  that  describes geometrical scaling  give the
 (one  standard deviation) allowed interval for $\sigma_{pp}^{\rm tot}$,
 as estimated in the  original Fly's Eye  publication.
 However it is  clear that  any  reasonable extrapolation  of the 
 collider data  (for the $B$--$\sigma_{pp}^{\rm tot}$) will result in the
 estimate of a higher  central value for the $pp$ cross and in larger
 uncertainty. Nominally the  prediction of Donnachie and Landshoff for
 $\sigma_{pp}^{\rm tot}$ at $\sqrt{s}  = 30$~TeV is  one  standard
 deviation  below the Fly's Eye  measurement.

  It is important to notice that the 
 experimentally measured and published inelastic $p$--air cross
 section is only that part of the total cross section which belongs to
 particle production. 
 Following \cite{Gaisser87a} we write this cross section as
\begin{equation}
\sigma_{p-air}^{\rm prod} = \sigma_{p-air}^{\rm tot} -
\sigma_{p-air}^{\rm el} - \sigma_{p-air}^{\rm q-el}\ ,
\label{sig-inel}
\end{equation}
 where $\sigma_{p-air}^{\rm q-el}$ is the quasielastic
 $p$--air cross section corresponding to scattering
 processes where the nucleus gets excited without direct
 particle production. The Glauber formalism~\cite{Glauber70a}
 gives explicit expressions for all terms in Eq.~\ref{sig-inel}.
 Unfortunately, there is ambiguity in the literature about
 the designation of the production cross section.  It has also
 been called $\sigma_{p-air}^{\rm inel}$
 in experimental~\cite{Baltrusaitis84,Baltrusaitis85,Honda93} and
 theoretical~\cite{Gaisser87a} papers and it 
is also often referred to as absorptive 
cross section \cite{Nikolaev86a,Durand88a,Kopeliovich89a,Nikolaev93b}.
 In the hope of removing this confusion, we introduce the notation
 `prod' to represent the inelastic cross section in which at least
 one new hadron is produced in addition to nuclear fragments.

\section*{Uncertainties in the $p$--air cross section measurement}

  In addition to uncertainties in converting from $\sigma_{p-air}^{\rm prod}$
 to $\sigma_{pp}^{\rm tot}$, there are significant uncertainties in
 the determination of $\sigma_{p-air}^{\rm prod}$ itself. Both at
 Fly's Eye~\cite{Baltrusaitis84} and at Akeno~\cite{Honda93}, the
 approach is to look at the frequency of deeply penetrating showers
 and to assign a corresponding  attenuation length ($\Lambda$)  on
 the assumption that, for a given energy, the most deeply
 penetrating showers are initiated by protons.

  The Fly's Eye group measures the depth of maximum development ($X_{\rm max}$)
 distribution for air showers in a relatively narrow interval of
 $S_{\rm max}$, where $S_{\rm max}\,\propto\,E_0$ is the shower size at maximum.
 The tail of that distribution, well after its peak, is a measure of the
 depth of the first interaction convoluted with the intrinsic fluctuations
 in the shower development.

  The Akeno group selects deeply penetrating showers by cutting on
 showers with the highest size, $S$, at the observation level in narrow
 bins of the shower muon size $S_\mu$. The reason for this procedure is
 that $S_\mu$ is nearly proportional to the primary energy $E_0$.
 $\Lambda$ is then derived from the frequency of such showers at
 different zenith angles, i.e. from the decrease of the frequency
 with atmospheric depth, which is a different measure of
attenuation from that used in the Fly's Eye approach.

 The model--dependence then is compressed into a single parameter $a>1$
 in the relation
\begin{equation}
\Lambda\;=\;a\times\lambda_{p-air}\;=\;a\times
{14.5\,m_p\over\sigma_{p-air}^{\rm prod}}.
\end{equation}
 Here $\lambda_{p-air}$ is the interaction length of protons in air, which
 has a mean atomic mass of 14.5.  The effective value of $a$ for proton
 initiated showers depends on the pion inelastic cross section in air and
 on the inclusive cross sections in the proton and pion inelastic
 interactions~\cite{Gaisser82,Bellandi95a}.

  The Fly's Eye proton air cross section value of 540$\pm$50 mb is
 derived by fitting the tail of the $X_{\rm max}$ distribution to an
 exponential with a slope of $\Lambda$ = 70$\pm$6 g/cm$^2$ and then using
 $a \, \simeq \,$1.60, which is similar to the value calculated in
 Ref.~\cite{Ellsworth82a}. The $\Lambda$ values in Ref.~\cite{Ellsworth82a}
 are calculated by simulating air showers assuming different energy
 dependences of $\sigma_{p-air}^{\rm prod}$ and fitting the tails of the
 resulting $X_{\rm max}$ distributions. $\Lambda$ values are then compared
 to $\sigma_{p-air}^{\rm prod}$ at primary energy $E_0>$3$\times$10$^{17}$~eV. 
 The calculation was performed with an essentially $pp$ scaling interaction
 model~\cite{Hillas81a}.

  Models with even very modest scaling violation, that also account for
 the nuclear target effect yield smaller values of $a$. It is not possible
 to separate the effects of the energy dependence of the inelastic cross
 section from those of the scaling violation in the `one parameter'
 approach. The relevant parameter in $p$--air interactions is the rate
 of energy dissipation by the primary proton
 $K^{\rm inel}_{p-air}/\lambda_{p-air}$. The inelasticity coefficient 
 $K_{p-air}^{\rm inel}  \, = \, {{E_0 - \langle E_L \rangle}
 \over {E_0}}$, where $E_0$ is the primary proton energy in the lab
 system and
 $\langle E_L \rangle$ is the average lab system energy of the leading
 nucleon. The equally strong contribution of $\pi$--air collisions is
 even more difficult to quantify in simple terms. On the other hand $a$
 tends to saturate for very strong scaling violation models, because the
 nuclear target effects in such models are small. 

 The Akeno experiment uses
 calculations~\cite{Kasahara79} made also with a model implementing
 radial scaling. In Ref.~\cite{Bellandi95a} the results of Akeno  have been 
 reanalyzed  making use of an interaction model with scaling violations,
 resulting in the derivation of lower values for $a$ that used by
 the Akeno experiment.
 
  We have updated the calculations of Ref.~\cite{Ellsworth82a} to
 illustrate how different values of $\sigma_{p-air}^{\rm prod}$  can
 be extracted from the same measured value of $\Lambda$ depending
 on the inclusive cross sections of the interaction model. 
 The calculations were performed with three interaction models
 characterized by $K_{p-air}^{\rm inel}$: the scaling model of
 Hillas~\cite{Hillas81a}, SIBYLL~\cite{Fletcher94} and a SIBYLL--based
 model with significantly stronger scaling violation in $pp$ interactions
 (High--K). All three calculations use the same 
input $\sigma_{p-air}^{\rm prod}=520$~mb
($\lambda_{p-air} \, = \, 46 \, {\rm g/cm}^2 $) at
 $\sqrt{s}$ = 30 TeV. The resulting values of 
$a$  are given in Table~\ref{a-tab}.  The last column of the table
gives the values of $\sigma_{p-air}^{\rm prod}$ that would be inferred
from the Fly's Eye measurements if the corresponding value of
$a$ had been used.
 The effects of scaling violation on the shower attenuation rate
 used by the Akeno experiment~\cite{Honda93} are similar, although
 the numerical values of $a$ are somewhat different.

\begin{table}[htb]
\caption{\em Cross section values that can be extracted from the
 measured $\Lambda$ = 70$\pm$6 g/cm$^2$ with different interaction
 models. 
\label{a-tab}
 }
\medskip
\begin{tabular}{lcccc}
Model & $\langle K_{pp}^{\rm inel} \rangle$
 & $\langle K_{p-air}^{\rm inel} \rangle$
 & $a(\sqrt{s}$ = 30 TeV)  & $\sigma_{p-air}^{\rm prod}$, mb\\
\tableline
 Hillas   &  ---  &  0.50  & 1.47$\pm$0.05 & 504 \\
 SIBYLL   & 0.57  &  0.67  & 1.20$\pm$0.05 & 411 \\
 High--K & 0.64  &  0.74  & 1.12$\pm$0.05 & 384 \\
\end{tabular}
\end{table}

\section*{Discussion}

  Cosmic--ray experiments detect air showers that result from 
 interactions of particles with energy up to and exceeding
 10$^{11}$ GeV. Such observations have the potential to provide
 information about the growth of $\sigma_{p-air}^{\rm prod}$ 
 up to $\sqrt{s}\, \simeq \, 10^5$ GeV. The long lever
 arm would be helpful for discriminating among
 models that give nearly identical results
 at lower energy. Here we attempt to summarize the problems and
 complications involved in the measurement and interpretation
 of $\sigma_{p-air}^{\rm prod}$ in cosmic ray experiments. 
 
 The experimental shower sets are inevitably contaminated by
 showers initiated by heavier nuclei. Neglecting this
 contamination would result in  an overestimate of 
 $\sigma_{p-air}^{\rm prod}$. To minimize this contamination,
the Fly's Eye cross section was estimated by analyzing
 only the most  penetrating   showers,  that is a 
 subset of 20\% of  the entire  data sample, strongly enriched
 in protons. A subsequent analysis 
 \cite{Gaisser93}  found that the composition of primary cosmic rays
 may be very  heavy in the energy region  considered.  If so, the
contamination
 of heavy  primaries  could  be larger  than  what was 
 estimated in the original work  leading to 
 an overestimate of $\sigma_{p-air}^{\rm prod}$.

 The cross section estimates 
 in Ref.~\cite{Baltrusaitis84,Baltrusaitis85,Honda93}  were
 based on interaction models with scaling particle
 momentum distributions. 
 Models with scaling violations predict faster shower development 
 (e.g. smaller values of  $a$).  If such models were used they
would imply a  smaller $p$--air cross section (as illustrated
in Table~\ref{a-tab}).  In addition, such models could also be
consistent with a smaller fraction of heavy nuclei.
 If the shower development is described with a single parameter, 
 as done in the first generation cross section estimates,
 it is impossible to distinguish between the effects of the
 proton and pion cross sections and the inclusive distributions
 of the secondary particles.

Once $\sigma_{p-air}^{\rm prod}$ is determined, the Glauber formalism
can be used to infer $\sigma_{pp}^{\rm tot}$ with extrapolations
for $B(s)$ based on all available collider data. 
Previous analyses~\cite{Baltrusaitis84,Baltrusaitis85,Honda93}
used a parameterization based on data up through ISR energies
which fails to describe recent high energy measurements and leads to an
underestimation of $\sigma_{pp}^{\rm tot}$.
 
 Our basic conclusion is that cosmic-ray values of $\sigma_{pp}^{\rm
tot}$
 do not at present strongly constrain extrapolations of fits
 of this cross section up to collider energies.  With the prospect
 of much more precise experimental measurements forthcoming from the
 high--resolution Fly's Eye and other proposed
 experiments~\cite{NewExp} there is the potential for
 much better estimates of the proton--proton cross section. 
 Realizing this potential will depend also on the use of
 a new generation \cite{Knapp96a} of shower simulations based on
 interaction models that incorporate all the physics of minimum
 bias interactions up to collider energies and a correspondingly
 detailed treatment of nuclear effects. The corresponding analysis
 should involve a full Monte Carlo simulation of each experimental
 data set rather than characterizing the simulation with a single
 parameter.


\noindent
 {\bf Acknowledgements.} \\
One of the authors (PL) wishes to thank the Bartol
Research Institute for hospitality during the time this work was completed.
One of the authors (R.E.) is grateful to J.\
Ranft and S.\ Roesler for many discussions.


\end{document}